\documentstyle[prl,eqsecnum,aps]{revtex}

\begin{document}
\draft
\preprint{}
\widetext
\title{Mean-Field Description of Phase String Effect in the $t-J$ Model}
\author{ Z.Y. Weng, D.N. Sheng, and C.S. Ting }
\address{Texas Center for Superconductivity and Department of Physics\\
University of Houston, Houston, TX 77204-5506}
\maketitle
\begin{abstract}
A mean-field treatment of the phase string effect in the $t-J$
model is presented. Such a theory is able to unite the antiferromagnetic (AF) 
phase at half-filling and metallic phase at finite doping within a single theoretical framework. We find that the low-temperature occurrence of the AF 
long range ordering (AFLRO) at half-filling and superconducting condensation 
in metallic phase are all due to Bose condensations of spinons and holons, 
respectively, on the top of a spin background  
described by bosonic resonating-valence-bond (RVB) pairing.  The fact that both 
spinon and holon here are bosonic objects, as the result of the phase string 
effect, represents a crucial difference from the conventional slave-boson and 
slave-fermion approaches. This theory also
allows an underdoped metallic regime where the Bose condensation of spinons  
can still exist. Even though the AFLRO is gone here, such a regime corresponds 
to a microscopic charge inhomogeneity with short-ranged spin ordering. We 
discuss some characteristic experimental consequences for those different metallic regimes. A perspective on broader issues based on the phase string 
theory is also discussed.

\end{abstract}


\widetext

\section{INTRODUCTION}

The $t-J$ Hamiltonian is one of the simplest nontrivial model to describe how 
doped holes move on the AF spin background and is widely used to characterize 
the physics in the $CuO_2$ layers of cuprates. A tremendous effort has been contributed to the investigation of the $t-J$ model. The 
most popular approaches to the metallic phase are often based on the so-called 
slave-boson method\cite{zba} in which the degrees of freedom associated with 
spins are described in terms of {\it fermionic} description. 
There have been many proposals of mean-field ground states based on such a 
fermionic description of spins, ranging from the earlier fermionic
RVB states,\cite{anderson} gauge-theory description,\cite{gauge1,gauge2} 
$SU(2)$ formalism,\cite{su(2)} to possible 
fractional-statistics.\cite{laughlin,wilczek} However, there is an inherent
problem quite general to the fermionic description of spins: approaches
based on it usually fail in faithfully producing correct AF correlations 
especially at small doping. 

At half-filling, for example, the exact ground state is known to
satisfy the Marshall sign rule\cite{marshall} (for a bipartite
lattice) but a fermionic description of spins would show redundant signs: 
even exchanging two same spins will give rise to a 
sign change of wavefunction due to the fermionic statistics. Under 
strict enforcement of no double occupancy constraint, those unphysical signs
would not have any effect. But in mean-field approximations, this ``sign 
problem'' will always show up and cause serious problem like overall underestimate of AF correlations.

By contrast, the Marshall sign can be easily incorporated into a 
{\it bosonic} description of spin degrees of freedom, where no extra sign 
problem would be caused by the statistics of bosons. It is the reason 
contributing to the success of the bosonic RVB description\cite{liang,chen} and 
its mean-field version -- the Schwinger-boson mean-field approach\cite{aa} in
describing spin properties at half-filling. A variational wavefunction based on 
the bosonic RVB picture can produce\cite{liang,chen} an unrivaled
accurate ground-state 
energy ($-0.3344J$ per bond as compared to the exact numerical value of 
$-0.3346J$ per bond for the Heisenberg model), and a generalized 
description\cite{chen} can precisely provide not only the ground-state energy, staggered magnetization, but also spin excitation spectrum in the whole 
Brillouin zone. Therefore, a bosonic description of spins seems most natural at
half-filling.

It is thus tempting for one to use the bosonic description of spins as a 
starting point and try to get into the metallic phase by doping.  
The connection between antiferromagnetic and metallic phases is commonly
perceived important in the $t-J$ model, and is also believed by many to be
the key in search for the mechanism of superconductivity in cuprates. 
Unfortunately, the mean-field study in the Schwinger-boson-slave-fermion approach\cite{lee} of the $t-J$ model, which is based on the bosonic description
of spins and has been quite successful at half-filling,\cite{aa} soon meets 
problematic consequences once holes are introduced -- encountering the so-called 
spiral phase and its derivatives.\cite{ss,spiral} It seems that 
one could not avoid such a spiral instability so long as a perturbative approach 
is adopted.\cite{weng1} One of many problems with spiral phases involves an 
underestimated kinetic energy $\propto\delta^2 \frac{t^2}{J}$ at weak-doping 
$\delta \ll 1$, which is 
also accompanied by a very quick descent to ferromagnetic phase at slightly 
larger $\delta$.\cite{spiral} 

This implies that doped holes may have introduced some {\it singular} doping effect which has been mistreated in the mean-field approximations. Such a 
singular doping effect has been recently identified\cite{string,string1} by reexamining the motion of doped holes in the AF background. It has been found
that spin-mismatches caused by the hopping of doped holes cannot be
completely ``repaired'' through spin flips at low energy. Such a residual 
nonrepairable effect can be expressed by a path-dependent phase product known as 
phase string.\cite{string} Due to the phase string effect,
a hole slowly moves through a closed path will acquire a nontrivial Berry's 
phase. As this phase
string effect is very singular locally at a lattice constant scale, its  
topological effect can be easily lost if a conventional mean-field average is
involved --- a reason causing the aforementioned spiral instability.  

In order to handle such a singular phase string effect hidden in the 
conventional Schwinger-boson-slave-fermion scheme, a unitary transformation\cite{string1} has
been introduced to {\it regulate} the Hamiltonian such that the local 
singularity of the phase string (at the scale of one lattice constant) is 
``gauged away'', while its large distance topological consequence is explicitly
incorporated into the Hamiltonian. The resulting exact reformulation\cite{string1} of the $t-J$ model is 
believed to be more suitable for a perturbative treatment, in contrast to the
original slave-fermion formalism. The underlying physical 
implication is that the ``holon'' and ``spinon'' {\it defined} in the 
slave-fermion scheme\cite{lee,spiral} may not be really separable due to the
hidden phase string effect, but those in the new formalism may become truely
elementary excitations. In the one-dimensional (1D) case, correct 
Luttinger-liquid behaviors indeed can be reproduced\cite{string1} after a 
mean-field decoupling of the spin and charge degrees of freedom in this new 
scheme. 

In this paper, we develop a generalized mean-field-type theory based on 
this new formalism of the $t-J$ model in the two-dimensional (2D) case. This 
theory recovers the well-known Schwinger-boson mean-field state\cite{aa} at 
half-filling while predicts a metallic phase at finite doping without 
encountering any  
spiral instability. It offers a unified phase diagram for the $t-J$ model at 
small doping, in which an insulating AFLRO phase, an underdoped metallic phase with the phenomena of pseudo-gap and charge inhomogeneity, as well as a uniform
metallic phase with ``optimized'' superconducting transition temperature,  
are all natural consequences happening on a single spin background controlled 
by bosonic RVB order pairing. The phase string effect plays a crucial role
here to connect those different phases together within a single theoretical
framework. A short version of this work was published earlier.\cite{letter}
 
\section{MEAN-FIELD THEORY BASED ON PHASE STRING EFFECT}

In the standard slave-fermion formalism of the $t-J$ Hamiltonian, 
electron annihilation operator 
$c_{i\sigma}$ is written as
\begin{equation}\label{swb}
c_{i\sigma}=f^{\dagger}_ib_{i\sigma}(-\sigma)^i,
\end{equation}
in which $f^{\dagger}_i$ is  fermionic ``holon'' creation operator and 
$b_{i\sigma}$ is bosonic (Schwinger-boson) ``spinon'' annihilation 
operator, satisfying no double occupancy constraint
\begin{equation}\label{constraint}
f^{\dagger}_if_i+\sum_{\sigma}b^{\dagger}_{i\sigma}b_{i\sigma}=1.
\end{equation} 
The $t-J$ model, $H_{t-J}=H_t+H_J$, is composed of two terms:
the hopping term $H_t$ is given by
\begin{equation}\label{ehop}
H_t=-t\sum_{\langle ij\rangle \sigma }\left(\sigma\right)f^{\dagger}_if_j 
b_{j\sigma}^{\dagger}b_{i\sigma}
 + H.c.,   
\end{equation}
and the superexchange term is
\begin{equation}\label{exc}
H_J=-\frac J 2 \sum_{\langle ij\rangle \sigma \sigma'}
b_{i\sigma}^{\dagger}b_{j-\sigma}^{\dagger}b_{j-\sigma'}b_{i\sigma'}.
\end{equation}
Note that the staggered phase factor $(-\sigma)^i$ in (\ref{swb}) is introduced\cite{string1} to explicitly track the Marshall sign, which leads to the negative sign in (\ref{exc}). A sign $\sigma=\pm 1$ then
appears in the hopping term (\ref{ehop}) which is the origin of the phase 
string effect\cite{string,string1} mentioned in the Introduction. Due to such a 
sign, a hole moving from a site $a$ to an another site $b$ will acquire a 
sequence of signs, i.e., a phase string as shown in Fig. 1, which has been 
shown\cite{string,string1} to be nonrepairable by the spin flip process governed by $H_J$. It implies that the slave-fermion formalism of the 
$t-J$ model cannot be treated in a perturbative way in doped case.

\subsection{Phase string representation}

It has been shown that the above singular effect of
phase string can be regulated after a unitary transformation.\cite{string1}
The resulting new formalism is known as the phase string representation.
The hopping term $H_t$ in this new representation becomes\cite{string1} 
\begin{equation}\label{et}
H_t=-t\sum_{\langle ij\rangle \sigma}\left(e^{iA_{ij}^f}\right)h^{\dagger}_ih_j
\left(e^{i\sigma A_{ji}^h}\right)b_{j\sigma}^{\dagger}b_{i\sigma} + H.c.   
\end{equation}
and the superexchange term $H_J$ reads 
\begin{equation}\label{ej}
H_J=-\frac J 2 \sum_{\langle ij\rangle \sigma \sigma'}
\left(e^{i\sigma
A_{ij}^h}\right)b_{i\sigma}^{\dagger}b^{\dagger}_{j-\sigma}\left(e^{i\sigma' 
A_{ji}^h}\right)b_{j-\sigma'}b_{i\sigma'}.
\end{equation}
Note that the fermionic operator $f_i$ now is replaced by a {\it bosonic} holon
operator $h_i$ in this new formalism. So one 
role of the phase string effect is to turn holons from fermions into bosons. 
Both holon and spinon are now described by bosonic operators which still 
satisfy the following no double occupancy constraint
\begin{equation}\label{constraint2}
h^{\dagger}_ih_i+\sum_{\sigma}b^{\dagger}_{i\sigma}b_{i\sigma}=1.
\end{equation}

In this new formalism, the singular phase string effect, as represented by the 
sign $\sigma$ in the hopping term (\ref{ehop}) of the original slave-fermion 
representation, is ``gauged away'', but its topological effect is left and 
exactly tracked by lattice gauge fields $A_{ij}^f$ and 
$A_{ij}^h$. These fields are defined as follows
\begin{equation}\label{eaf}
A_{ij}^f\equiv A_{ij}^s-\phi_{ij}^0
\end{equation}
with 
\begin{equation}\label{eas}
A_{ij}^s=\frac 1 2 \sum_{l\neq i, j}\left[\theta_i(l)-\theta_j(l)
\right]\left(\sum_{\sigma}\sigma n_{l\sigma}^b\right),
\end{equation}
\begin{equation}\label{epip}
\phi_{ij}^0=\frac 1 2 \sum_{l\neq i, j}\left[\theta_i(l)-\theta_j(l)
\right],
\end{equation}
and
\begin{equation}\label{eah}
A_{ij}^h=\frac 1 2 \sum_{l\neq i, j}\left[\theta_i(l)-\theta_j(l)\right]n_l^h.
\end{equation}
Here $n_{l\sigma}^b$ and $n_l^h$ are spinon and holon number operators, 
respectively. $\theta_i(l)$ is defined as an angle
\begin{equation}\label{angle}
\theta_i(l)=\mbox{Im ln $(z_i-z_l)$}
\end{equation}
with $z_i=x_i+iy_i$ representing the complex coordinate of a lattice site $i$.

The physical meaning of $A_{ij}^s$ and $A_{ij}^h$ have been discussed in 
Ref.\onlinecite{string1}: $A_{ij}^s$ and $A_{ij}^h$ describe 
quantized flux tubes bound to spinons and holons, respectively (Fig. 2 illustrates the case for $A_{ij}^h$). 
Furthermore, the field $\phi_{ij}^0$ describes a uniform flux threading through 
the 2D plane with a strength $\pi$ per plaquette: $\sum_{\Box}\phi^0_{ij}=\pm 
\pi$. It is also noted that a $\pi$-flux neutral topological excitation has been 
previously discussed\cite{ng} in the {\it pure} Heisenberg model, which 
resembles a quantized flux line in the mixed phase of a BCS superconductor. Here
the flux quanta are bound to the doped holes due to the phase string effect.

Electron operator in this representation becomes 
\cite{string1}
\begin{equation}\label{emutual}
c_{i\sigma}=h_i^{\dagger}b_{i\sigma}(-\sigma)^ie^{i\Theta_{i\sigma}^{string}}.
\end{equation}
Here the nonlocal phase factor $e^{i\Theta_{i\sigma}^{string}}$ precisely keeps
the track of the singular part of the phase string effect and is defined by
\begin{equation}\label{theta}
\Theta_{i\sigma}^{string}\equiv \frac 1 2 [\Phi^b_i-\sigma \Phi_i^h],
\end{equation}
with
\begin{equation}\label{phib}
\Phi^{b}_i\equiv \sum_{l\neq i} 
\theta_i(l)\left(\sum_{\alpha}\alpha n_{l\alpha}^b -1\right), 
\end{equation} 
and
\begin{equation}\label{phih}
\Phi_i^h\equiv \sum_{l\neq i}\theta_i(l) n_l^h. 
\end{equation}

\subsection{Mean-field approximation}
For the sake of clarity, in the following we first consider the
superexchange term $H_J$ and then include the hopping term $H_t$, due to 
different natures represented by them.

\subsubsection{Generalized mean-field treatment of $H_J$}

At half-filling, the mean-field theory based on the bosonic RVB picture is known
as the Schwinger-boson mean-field theory which was first introduced by Arovas and Auerbach\cite{aa}. Such a mean-field is characterized by a bosonic RVB order
parameter 
\begin{equation}\label{order0}
\Delta^s=\sum_{\sigma}\langle b_{i\sigma}b_{j-\sigma}\rangle
\end{equation}
for the nearest-neighbor sites $i$ and $j$ [$i= nn(j)$].
 
The present formalism only differs from the Schwinger-boson, slave-fermion 
formalism in doped case, where a gauge field $A_{ij}^h$ emerges. Since spinons 
are subject to this gauge-field $A_{ij}^h$ in $H_J$, it is natural to 
incorporating the link variable $e^{-i\sigma A_{ij}^h}$ into the order parameter 
(\ref{order0}). Namely,
\begin{equation}\label{order}
\Delta_{ij}^s=\sum_{\sigma}\left\langle e^{-i\sigma A^h_{ij}}b_{i\sigma}b_{j-\sigma}\right\rangle.
\end{equation}
$\Delta_{ij}^s$ defined here is then ``gauge-invariant'' under an ``internal'' 
gauge transformation: $A_{ij}^h\rightarrow A^h_{ij}+\theta_i-\theta_j$, and 
$b_{i\sigma}\rightarrow b_{i\sigma}e^{i\sigma \theta_i}$. 

Based on such an order parameter, one may write down the mean-field version of $H_J$ in (\ref{ej}) in a standard procedure
\begin{equation}\label{hjs}
H_s^{J}=-\frac{J}{2}\sum_{\langle ij \rangle \sigma}\left(\Delta^s_{ij}
\right)^*e^{-i\sigma A_{ij}^h}b_{i\sigma}b_{j-\sigma} + H.c. +
\frac J 2 \sum_{\langle ij \rangle} |\Delta^s_{ij}|^2+ 
\lambda \left(\sum_{i\sigma}b^{\dagger}_{\sigma}b_{i\sigma}-
(1-\delta)N \right),
\end{equation}
where the last term with a Lagrangian multiplier $\lambda$ is introduced to
enforce the condition of total spinon number, with $N$ denoting the total
lattice number and $\delta$ doping concentration. In order to diagonalize
$H_s^J$, we introduce the following Bogoliubov transformation
\begin{equation}\label{bogo}
b_{i\sigma}=\sum_m\left( u_{m\sigma}(i)\gamma_{m\sigma}-v_{m\sigma}(i)
\gamma^{\dagger}_{m-\sigma}\right),
\end{equation}
and seek the solution
\begin{equation}\label{eom1}
\left[H_s^J, \gamma_{m\sigma}^{\dagger}\right]=E_m \gamma_{m\sigma}^{\dagger},
\end{equation}
\begin{equation}\label{eom2}
\left[H_s^J, \gamma_{m\sigma}\right]=-E_m \gamma_{m\sigma}. 
\end{equation}
Here $\gamma_{m\sigma}$ and $\gamma_{m\sigma}^{\dagger}$ are bosonic
annihilation and creation operators, respectively, for an
eigenstate with quantum number $m$ and spin $\sigma$. In terms of bosonic 
commutation relations, one easily finds that $u_{m\sigma}(i)$ and 
$v_{m\sigma}(i)$ satisfy 
\begin{equation}\label{uu}
\sum_m\left[u_{m\sigma}(i)u_{m\sigma}^*(j)-v_{m\sigma}(i)v^*_{m\sigma}(j)\right]
=\delta_{ij}\ \  ,
\end{equation}
and
\begin{equation}\label{vv}
\sum_m\left[u_{m\sigma}(i)v_{m-\sigma}(j)-v_{m\sigma}(i)u_{m-\sigma}(j)\right]=0.
\end{equation}
According to (\ref{hjs}), we have
\begin{equation}\label{blinear}
\left[H_s^J, b_{i\sigma}\right]=\frac{J}{2}\sum_{j=nn(i)}\Delta_{ij}^se^{-i\sigma A_{ji}^h}b^{\dagger}_{j-\sigma}-\lambda b_{i\sigma}.
\end{equation}
Then by using (\ref{bogo}), (\ref{eom1}) and 
(\ref{eom2}), it is straightforward to derive the
following relations from (\ref{blinear}): 
\begin{eqnarray}
\label{u&v1}
- E_m u_{m\sigma}(i)& =-& \frac J 2 \sum_{j=nn(i)}\Delta^s_{ij}e^{-i\sigma A_{ji}^h}v^*_{m-\sigma}(j) -\lambda u_{m\sigma}(i), \\
\label{u&v2}
- E_m v_{m\sigma}(i)& =& \frac J 2 \sum_{j=nn(i)}\Delta^s_{ij}e^{-i\sigma A_{ji}^h}u^*_{m-\sigma}(j)   +\lambda v_{m\sigma}(i).
\end{eqnarray}
We can further express $u_{m\sigma}(i)$ and $v_{m\sigma}(i)$ by the ``one-particle'' wavefunction $w_{m\sigma}(i)$ as follows 
\begin{eqnarray} 
u_{m\sigma}(i)& = & u_m w_{m\sigma}(i), \\ 
v_{m\sigma}(i)&= &  v_mw_{m\sigma}(i), 
\end{eqnarray}
where $u_m$ and $v_m$ will be taken to be real and satisfy 
\begin{equation}\label{uv} 
u_m^2-v_m^2=1. 
\end{equation}
Then $w_{m\sigma}(i)$ is normalized following from (\ref{uv}):
\begin{equation}\label{orthog} 
\sum_m w_{m\sigma}(i)w^*_{m\sigma}(j)=\delta_{ij}. 
\end{equation}
Equations (\ref{u&v1}) and (\ref{u&v2}) then reduce to an  eigen-equation for 
the one-particle wavefunction $w_{m\sigma}$:
\begin{equation}\label{ew}
\xi_m w_{m\sigma}(i)=-\frac J 2 \sum_{j=nn(i)} \Delta^s_{ij}e^{-i\sigma A_{ji}^h}w^*_{m-\sigma}(j).
\end{equation}
The eigenvalue $\xi_m$ in (\ref{ew})
is related to $E_m$ and $u_m$ and $v_m$ as follows  
\begin{equation}\label{xi}
\xi_m=-(E_m-\lambda_m)\frac{u_m}{v_m}=(E_m+\lambda_m)\frac{v_m}{u_m}.
\end{equation}
Here $\lambda_m$ is the same as the Lagrangian multiplier $\lambda$, but we write it in a general form because later it will be modified once the hopping term is introduced. In terms of (\ref{xi}) and (\ref{uv}), one obtains 
\begin{equation}\label{espinon}
E_m=\sqrt{\lambda_m^2-\xi_m^2},
\end{equation}
and
\begin{equation}\label{u}
|u_m|  =\frac 1{\sqrt{2}}\left(\frac{\lambda_m}{E_m}+ 1\right)^{1/2},
\end{equation}
\begin{equation}\label{v}
|v_m|  =\frac 1{\sqrt{2}}\left(\frac{\lambda_m}{E_m}- 1\right)^{1/2}.
\end{equation}
The signs of $u_m$ and $v_m$ are determined up to  $sgn (v_m/u_m)=sgn(\xi_m)$. 
As a convention we will always choose $u_m=|u_m|$ and $v_m=|v_m|sgn(\xi_m)$. 

Thus $H_s^J$ is diagonalized as $H_s^J=\sum_{m\sigma}E_m\gamma_{m\sigma}^{\dagger}\gamma_{m\sigma}$ + const.  
according to (\ref{eom1}) and (\ref{eom2}).
The order parameter $\Delta_{ij}^s $ can be self-consistently determined by the definition (\ref{order}) as
\begin{equation}\label{delta}
\Delta^s_{ij} = \sum_{m\sigma}e^{-i\sigma A_{ij}^h}w_{m\sigma}(i)w_{m-\sigma}(j)
(-u_mv_m)\left[1+\sum_{\alpha}\langle 
\gamma_{m\alpha}^{\dagger}\gamma_{m\alpha}\rangle\right].
\end{equation}
In the following we will always consider the solution of a real order parameter $\Delta_{ij}^s$. In this case,  
it can be checked self-consistently that $w_{m\sigma}=w^*_{m-\sigma}$ according to (\ref{ew})
and $\left(\Delta_{ij}^s\right)^*=\Delta_{ij}^s$ in terms of (\ref{delta}).
The order-parameter equation may be further simplified if one multiplies (\ref{delta}) by $\left(\Delta_{ij}^s\right)^*$ and sums over $\langle ij \rangle$ with using (\ref{ew}):
\begin{equation}\label{d}
\sum_{\langle ij \rangle}\left|\Delta^s_{ij}\right|^2=\sum_m\frac 
{\xi_m^2}{JE_m} \coth \frac {\beta E_m}{2},
\end{equation}
with $\langle \gamma_{m\sigma}^{\dagger}\gamma_{m\sigma}\rangle=1/(e^{\beta E_m}-1)$ and $\beta\equiv 1/k_B T$.
Finally, the condition
\begin{equation}\label{number}
\left\langle\sum_{i\sigma}b^{\dagger}_{i\sigma}b_{i\sigma}\right\rangle=(1-\delta)N,
\end{equation}
which is enforced by the Lagrangian multiplier in (\ref{hjs}) and can be
rewritten as
\begin{equation}\label{lambda}
2-\delta=\frac 1 N \sum_{m\neq 0} \frac {\lambda_m}{E_m}\coth \frac{\beta E_m}{2}+n_{BC}^b,
\end{equation} 
where $n_{BC}^b$ is introduced to describe the contribution from $E_m=0$ state 
(denoted by $m=0$) when the Bose condensation of spinons occurs.\cite{bc} 
In comparison with the zero-doping Schwinger-boson mean-field theory, the above Bogoliubov-de Gennes scheme at finite doping mainly differs in the one-particle 
eigenequation (\ref{ew}) (and the resulting  energy spectrum  $\xi_m$ and wavefunction $w_{m\sigma}(i)$). One may simply recover the Schwinger-boson 
mean-field results by setting $A_{ij}^h=0$ in (\ref{ew})
and obtaining $\xi_k=-J\Delta_0^s(\cos k_xa+\cos k_y a)$ and the Bloch wavefundtion $w_{k\sigma}=\frac{1}{\sqrt{N}}e^{i\sigma{\bf k}\cdot {\bf r}_i}$
in the no-hole case. 

So the doping effect has entered the above mean-field theory in two ways:
one is through the particle number condition in (\ref{number}); the
other is through the gauge field $A_{ij}^h$. For the mean-field theory to work, 
$A_{ij}^h$ has been implicitly assumed as a time-independent field. But  
with holons moving around, $A_{ij}^h$ will usually gain a
dynamic effect. To see this, let us consider the following gauge-invariant
quantity 
\begin{equation}
\sum_C A_{ij}^h=\pi\sum_{l\in C} n_l^h,
\end{equation}
where $C$ is an arbitrary counter-clockwise closed path. If one redefines
\begin{equation}
n_l^h=\delta +\delta n_l^h,
\end{equation}
with $\delta n^h_l=n^h_l-\delta$, then correspondingly 
\begin{equation}\label{ah}
A_{ij}^h=\bar{A}_{ij}^h+\delta A_{ij}^h,
\end{equation}
where 
\begin{equation}\label{ahb}
\sum_C \bar{A}_{ij}^h =\pi \delta \frac{S_c}{a^2}\equiv 
\bar{\phi}\frac{S_c}{a^2},
\end{equation}
($S_c$ denotes the area of a loop $C$ and $a$ is the lattice constant) and
\begin{equation}\label{ahd}
\sum_C\delta A_{ij}^h=\pi \sum_{l\in C}\delta n^h_l.
\end{equation}
So the dynamics of $A_{ij}^h$ is determined by the fluctuations of the
density of holons on lattice. But since spinons and holons here are treated as 
independent degrees of freedom, one may neglect the dynamical effect of $\delta A_{ij}^h$ on spinon part at the mean-field level and replace it by some
random flux fluctuations with a strength per plaquette equal to $\delta\phi $ (One may estimate 
$\delta\phi \approx \pi\sqrt{(\delta n^h)^2}$). This can be justified at low
temperature when a Bose condensation of holons (which corresponds to a 
superconducting condensation as shown later) occurs, where $\delta A_{ij}^h$ is
expected to be substantially suppressed. On the other hand, however, in the high
temperature phase where the motion of holons is much less coherent, the 
fluctuation effect of $\delta A_{ij}^h$ can dominate over $\bar{A}^h_{ij}$ and
the separation of the latter from $A^h_{ij}$ then becomes meaningless. In this
case, one may approximately describe the effect of $A^h_{ij}$ as a collection of 
randomly distributed $\pi$ flux quanta with the number equal to that of holons. 
In both limits, the dynamics of $A_{ij}^h$ may be neglected. 

\subsubsection{Including the hopping term $H_t$}

First of all, we note that the wavefuntion $w_{m\sigma}(i)$ as the solution of 
the linear equation (\ref{ew}) is not unique, and it can be always multiplied by 
an arbitrary global phase factor $e^{i\sigma\chi_m}$: i.e.,
\begin{equation}\label{ps}
w_{m\sigma}(i)\rightarrow e^{i\sigma\chi_m} w_{m\sigma}(i),
\end{equation}
without changing the order parameter $\Delta^s_{ij}$ and the mean-field state.
Correspondingly the Bogoliubov transformation can be generally rewritten as
\begin{equation}\label{bogo2}
b_{i\sigma}=\sum_m\left( u_{m}\gamma_{m\sigma}-v_{m} \gamma^{\dagger}_{m-\sigma}\right) e^{i\sigma\chi_m}w_{m\sigma}(i).
\end{equation}
In particular, $e^{i\sigma\chi_m}$ can depend on the {\it holon}
configurations because the Hilbert space of $b_{i\sigma}$ is only well-defined 
at each given holon configuration due to the no-double-occupancy constraint. 
The hopping term will mix the Hilbert space of $b_{i\sigma}$ at different holon 
configurations, and such a freedom in phase choice can be fixed by optimizing 
the hopping integral of holons below. 

Now consider the hopping term $H_t$ in (\ref{et}). By using the Bogoliubov 
expression for spinon operators, a straightforward calculation gives 
\begin{equation}\label{bb}
\left\langle\sum_{\sigma} e^{i\sigma A^h_{ji}}b^{\dagger}_{j\sigma}b_{i\sigma} \right\rangle =
\sum_{m\sigma} e^{i\sigma A^h_{ji}}w^*_{m\sigma}(j)w_{m\sigma}(i)e^{-i\sigma\Delta \chi_m}\left[v^2_m+(u_m^2+v^2_m)\langle\gamma^{\dagger}_{m\sigma}\gamma_{m\sigma}\rangle\right].
\end{equation}
Note that $\Delta \chi_m$ in the above expression denotes the difference of 
$\chi_m$ before and after the holon changes the position. If one simply chooses $\Delta\chi_m=0$, namely, the phases of $w_{m\sigma}(i)$ to be the same for all
hole configurations, then the right-hand side of (\ref{bb}) vanishes for 
the nearest-neighboring $i$ and $j$. This can be verified by noting that $w_{\bar{m}\sigma}(i)\equiv (-1)^iw_{m\sigma}$ is also a solution of (\ref{ew}) with an eigenvalue ${\xi}_{\bar{m}}=-\xi_m$ and  the cancellation
in (\ref{bb}) stems from the fact that those  quantities like $u_m^2$ and $v_m^2$ only depend on $E_m$ in (\ref{bb}) which is symmetric under $\xi_m\rightarrow -\xi_m$. But such a cancellation is removable by a simple 
choice of the phase shift in $e^{i\sigma\chi_m}$  at different holon 
configurations when each time a holon changes sublattice sites: 
\begin{equation}\label{wphase}
e^{i\sigma\chi_m}\rightarrow \left[-sgn(\xi_m)\right] \times e^{i\sigma\chi_m},
\end{equation}
or 
\begin{equation}
e^{-i\sigma\Delta\chi_m}=-sgn(\xi_m).
\end{equation}
Then,  one finds
\begin{equation}\label{B0}
B_0\equiv\frac{1}{2N} \sum_{\langle ij\rangle\sigma}\left\langle
e^{i\sigma A^h_{ji}}b^{\dagger}_{j\sigma}b_{i\sigma} \right\rangle=\frac{1}{2N}
\sum_{m\sigma}B^0_m\left[v^2_m+(u_m^2+v^2_m)\langle\gamma^{\dagger}_{m\sigma}\gamma_{m\sigma}\rangle\right],
\end{equation}
where $B^0_m\equiv \sum_{\langle ij\rangle}e^{i\sigma A^h_{ji}}w^*_{m\sigma}(j)w_{m\sigma}(i)\left[-sgn(\xi_m)\right]$ is given by
\begin{eqnarray}
B^0_m = & -sgn (\xi_m) \sum_{\langle ij\rangle}e^{i\sigma A^h_{ji}}w^*_{m\sigma}(j)w_{m\sigma}(i) \nonumber\\
=& sgn(\xi_m) \frac{{\xi}_m}{2J_s}\sum_iw^*_{m\sigma}(i)w_{m\sigma}(i) \ \ \ 
\nonumber\\
=& \frac{|{\xi}_m|}{2J_s}\ \  >0 \ \ \ \ \ \ \ \ . 
\ \ \ \ \ \ \ \ \ \ \ \ \ \ \ \ \ \  \ 
\end{eqnarray}
In obtaining the second line above, we have used (\ref{ew}) with $\Delta_{ij}^s=
\Delta^s$ and $J_s\equiv \frac{1}{2}\Delta^s J$. 

Holons thus acquire a finite hopping integral without introducing any 
{\it extra} order parameter. The effective holon Hamiltonian is
given by:
\begin{equation}\label{hh}
H_h=-t_h\sum_{\langle ij\rangle}e^{i A^f_{ij}}h^{\dagger}_ih_j + H.c.,
\end{equation} 
which is derived from $H_t$ with an effective hopping integral
\begin{equation}
t_h\equiv t B_0.
\end{equation}
It is important to note that such a {\it finite} kinetic energy ($t_h\sim 
t$) that each holon has gained on the present mean-field spin background cannot 
be similarly realized in the slave-fermion-Schwinger-boson scheme, exactly due 
to the hidden phase-string effect: the sign $\sigma=\pm 1$ in (\ref{ehop}) will
always lead to $t_h=0$ and make  a spiral 
twist (a new order parameter in favor of hole hopping) necessary in any local mean-field  treatment. 

Finally, to be consistent, the hopping effect {\it on} spinons
is obtained from $H_t$ as:
\begin{equation}\label{hst}
H^t_s=-J_h \sum_{\langle ij\rangle\sigma} e^{i\sigma A^h_{ji}}b^{\dagger}_{j\sigma}b_{i\sigma} + H.c. +4J_hB_0N,
\end{equation}
in which $J_h\equiv \langle e^{iA^f_{ij}}h^{\dagger}_ih_j\rangle t\propto \delta 
t$ measures the strength of hopping effect on the spinon part. (The constant in 
(\ref{hst}) is introduced such that $\langle H_s^t\rangle=0$.) Hence the 
total Hamiltonian describing spinon degrees of freedom is composed of two terms 
\begin{equation}
H_s= H_s^J + H_s^t, 
\end{equation} 
where $H_s^J$ in (\ref{hjs}) has been diagonalized at the mean-field level 
before. $H_s^t$ can be expressed in terms of (\ref{bogo2}) by noting that $b^{\dagger}_{j\sigma}$ and $b_{i\sigma}$ in (\ref{hst}) should differ by a 
phase shift (\ref{wphase}) as a holon switches sublattice sites. It
then gives
\begin{equation}
H_s^t=-J_h\sum_{m\sigma}B^0_m (u_m\gamma_{m\sigma}^{\dagger}-v_m\gamma_{m-\sigma})(u_m\gamma_{m\sigma}-v_m\gamma^{\dagger}_{m-\sigma}) + H.c. +4J_hB_0N,
\end{equation}
after using the orthogonal condition $\sum_i w^*_{m\sigma}(i)w_{m'\sigma}(i)=\delta_{m,m'}$ as well as (\ref{ew}).  
On the other hand, one has $H_s^J=\sum_{m\sigma}E_m\gamma^{\dagger}_{m\sigma}\gamma_{m\sigma}+$ const. 
(note that $\lambda_m=\lambda$ inside $E_m$ here). Then $H_s$ can be 
diagonalized in a 
straightforward way. If we are still to use $E_m$ to denote th
spinon spectrum for the sake of compactness, then $H_s$ can be finally written 
as
\begin{equation}\label{hs}
H_s=\sum_{m\sigma} E_m \gamma_{m\sigma}^{\dagger}\gamma_{m\sigma} + E_0^s,
\end{equation}
where 
\begin{equation}
E_0^s=-2\sum_m E_m n(E_m) - \frac J 2 \sum_{\langle ij\rangle} 
|\Delta^s_{ij}|^2.
\end{equation}
But here the spinon spectrum 
\begin{equation}\label{espinon2}
E_m=\sqrt{\lambda_m^2-{\xi}_m^2} 
\end{equation}
is different from the previous one obtained in previous section by a correction 
to $\lambda_m$ due to the hopping effect: 
\begin{equation}\label{lambda2} 
\lambda_m=\lambda-\frac{J_h}{J_s}|{\xi}_m|.
\end{equation}

Therefore, the hopping effect on the spinon part is {\it solely} represented by 
a shift from $\lambda$ to $\lambda_m$ in (\ref{lambda2}). The Bogoliubov transformation 
(\ref{lambda2}) remains unchanged and so do $u_m$  and $v_m$ defined in (\ref{u}) and (\ref{v}), so long as the renormalized $\lambda_m$ is used.
The Lagrangian mutiplier $\lambda$ in (\ref{lambda2}) is still determined by
(\ref{lambda}) where both $\lambda_m$ and $E_m$ should be replaced by the
renormalized ones in (\ref{espinon2}) and (\ref{lambda2}). Finally, the 
self-consistent equation (\ref{d}) for the RVB pairing order parameter maybe 
rewritten as
\begin{eqnarray}\label{ds}
\Delta^s=\frac{1-2\delta}{4N}\sum_{m}\frac{{\xi}_m^2}{J_s{E}_m}\coth \frac { {\beta}{E}_m}{2}.
\end{eqnarray}
(Note that if the Bose condensation occurs, one may separate the contribution
from the condensation part on the right-hand-side by $\Delta_{BC}^s\equiv (1-
2\delta)|\xi_0|^2n_{BC}^b/4J_s\lambda_0$). In obtaining (\ref{ds}), we 
have used an approximate relation $1/(2N)\sum_{\langle ij \rangle} \left\langle(\Delta_{ij}^s)^2\right\rangle_h\approx {\Delta^s}^2/(1-2\delta)$ at
$\delta \ll 0.5$ limit. Such a relation can be obtained by assuming  
$\Delta_{ij}^s=\Delta^s_1$ when  $i$ and $j$ belongs to occupied sites and
$\Delta^s_{ij}=0$ if $i$ or $j$ is at hole site and by noting that each hole 
accounts for $\Delta^s_{ij}=0$ at four adjacent bonds at dilute-hole limit which 
leads to $\Delta^s=(1-2\delta)\Delta^s_1$.

\section{PHASE DIAGRAM}

\subsection{Unified bosonic RVB phase}

Our mean-field theory has been constructed based on a single 
bosonic RVB order parameter $\Delta^s$. Such an order parameter controls 
short-range
spin-spin correlations in {\it both} undoped and doped regime. Fig. 3 shows a 
typical region of $\Delta^s\neq 0$ obtained by solving the
mean-field equations which has been briefly discussed in 
Ref.\onlinecite{letter}. It obviously covers the whole experimentally interested 
temperature (from $T=0$ to $T\sim 0.5-0.9 J/k_B$) and doping (from $\delta=0$ to 
$\delta>0.3$) regime. Several low-temperature regions {\it within} this phase as 
marked in Fig. 3, including the superconducting phase, will be discussed in the 
following sections. The normal-state within this phase will
correspond to a ``strange metal'' phase, where magnetic 
and transport properties are expected to be different from conventional metals. 
It is noted that in the bosonic RVB description of spin 
degrees of freedom, the order parameter $\Delta^s$ does not directly correspond 
to an energy gap, in contrast to the fermionic RVB theory\cite{zba} (the latter
is similar to the BCS theory in mathematical structure). Also note that the 
crossover from $\Delta^s\neq 0$ phase to $\Delta^s=0$ phase at high temperature 
is similar to the 
half-filling case\cite{aa} which does not correspond to a real phase transition. 

In obtaining $\Delta^s$ in 
Fig. 3 by solving (\ref{ds}), one also needs to determine the spectrum ${\xi}_m$
from (\ref{ew}) and decide the chemical potential $\lambda$ in terms of 
(\ref{lambda}). We have chosen the parameter $J_h=\delta J$ (which corresponds
to $t\sim J$) and solved ${\xi}_m$ under $A^h_{ij}=\bar{A}^h_{ij}$, but other 
choices of $J_h$ as well as including the fluctuating part $\delta A_{ij}^h$ do 
not change significantly the range covered by $\Delta^s\neq 0$. The effect of
$\delta A_{ij}^h$ will be the subject of discussion in the next section, and we
will always use the same $J_h$ below. 

Although $\Delta^s\neq 0$ practically covers the whole doping regime, at a 
larger doping concentration, this mean-field theory may no longer be
energetically favorable due to the competition between the 
hopping and superexchange energies. Actually, the phase string effect itself is 
an indication that the bosonic description of spins leads to frustration of
the motion of doped holes, and vise versa. With the increase of doping 
concentration, one possibility is that eventually a statistical-transmutation 
may occur to effectively turn bosonic spinons into fermionic ones as
to be discussed in Sec. IV. Beyond such a 
point, the present mean-field theory will break down, which may determine a 
crossover to the so-called overdoped regime. We will explore this issue 
elsewhere. 

\subsection{Bose condensation of spinons: AF ordering vs. phase separation}

\subsubsection{AFLRO and insulation phase}

At half-filling, the spinon spectrum $E_m$ is known to be { 
gapless} at zero doping and zero temperature which ensures a 
Bose condensation\cite{bc} of spinons. Such a Bose condensation of spinons, 
as represented by $n_{BC}^b\neq 0$ in (\ref{lambda}), describes a long-range AF 
spin ordering.\cite{bc} The Bose condensation or long-range AF order can be 
sustained up to a finite temperature $T_N> 0$ if
the three-dimensional effect (interlayer coupling) is included. In the 
following, we consider how this AFLRO picture evolves at finite doping. 

Based on the expression (\ref{emutual}), spin operators, $S^z_i$ and 
$S^{\pm}_i$, can be easily written down in terms of
spinon operator $b_{i\sigma}$ after using the constraint (\ref{constraint2}):
\begin{equation}\label{sz}
S^z_i=\frac 1 2 \sum_{\sigma}\sigma b^{\dagger}_{i\sigma}b_{i\sigma},
\end{equation}
and
\begin{equation}\label{s+}
S^{+}_i=b^{\dagger}_{i\uparrow}b_{i\downarrow}(-1)^ie^{i\Phi_i^h}
\end{equation}
and $S^-_i=(S^+_i)^{\dagger}$.

At $\delta=0$, one has $\Phi_i^h=0$, and the Bose condensation leads to 
\begin{equation}
\langle S^+_i\rangle\propto 
(-1)^i,
\end{equation}
i.e., an AFLRO. But at $\delta\neq 0$, even when the spinons are 
Bose-condensed, $\langle S^+_i\rangle$ should generally vanish due to the fact
that 
\begin{equation}\label{phih0}
\langle e^{i\Phi_i^h}\rangle=0.
\end{equation}
The proof here is straightforward. Note that in the definition of 
$\Phi_i^h$ in (\ref{phih}), the angle $\theta_i(l)$ [(\ref{angle})] can be 
transformed as 
\begin{equation}
\theta_i(l)\rightarrow \theta_i(l)+\phi
\end{equation}
for an arbitrary $\phi$ {\it without changing} $A_{ij}^h$, $A_{ij}^f$, and thus the Hamiltonian. But $e^{i\Phi_i^h}$ [(\ref{phih})] changes accordingly 
\begin{equation}
e^{i\Phi_i^h}\rightarrow e^{i\Phi_i^h}\times e^{i\phi N^h},
\end{equation}
Here $N^h$ is the total holon number.  Thus the average of such a
phase must vanish at finite doping as given in (\ref{phih0}).  Since $\Phi^h_i$ describes vortices centered at holons, it is like a free-vortex phase as holons
move around freely in metallic phase, which resembles 
a disordered phase in a Kosterlitz-Thouless-type transition. 

Only in the case that holons are localized (i.e., in insulating phase) such
that the phase string effect is ineffective, the AFLRO may be 
recovered. In this insulating phase, holons are perceived by spinons as 
localized vortices like in the mixed state of a type-II superconductor, and by 
forming ``supercurrents'' to screen those vortices, $\Phi^h_i$ in $S^+_i$ can
be effectively canceled out by the opposite vorticities generated from spinons.
After all, the phase string effect is no longer important if the motion of 
holons is limited. We do not know if the localization of holons can happens intrinsically or has to be under some external factors like impurity effect. But
we expect such an insulating phase to exist only at a very dilute density of 
holons at the expense of latter's kinetic energy. 

\subsubsection{Bose condensation of spinons in metallic phase: underdoping}

We have shown that the AFLRO must be absent in metallic phase. One may 
naturally wonder if the Bose condensation of spinons can still persist into
metallic phase, and if it does, then what is its physical meaning? 

To answer these questions, let us first to inspect how
the spinon spectrum $E_m$ is modified by doping. In Fig. 4, we compare the spinon density 
of states $\rho_s(E)$ at $\delta=1/7\approx 0.143$ (solid curve) with the 
$\delta=0$ case (diamond curve in the insert). Here $\rho_s(E)$ is defined by 
\begin{equation}\label{rhos}
\rho_s(E)=\frac 1 N \sum_m\delta(E-E_m).
\end{equation}
One notices that a unique peak-structure is clearly exhibited at $\delta=0.143$. 
This can be easily understood by noting that the spinon
spectrum $E_m$ is basically determined by ${\xi}_m$ which, as the solution
of (\ref{lambda}), has a Hofstadter structure (or the Landau levels in the 
continuum limit) due to a uniform flux ($\bar{\phi}=\pi\delta$ per plaquette) 
represented by the vector potential 
$\bar{A}_{ij}^h$ threading through the square lattice. The broadening of the 
solid curve in Fig. 4 is due to the redistribution of eigenstates under the
fluctuating flux $\delta A_{ij}^h$, which is treated as a random flux (in white
noise limit) here 
with a maximum strength chosen at $\delta\phi=0.3 \bar{\phi}$ per plaquette.
By contrast, the dashed curve marks the positions of sharp peaks of density of 
states in the limit of $\delta\phi=0$. 

We find that the Bose condensation of spinons can still occur at $\delta=0.143$ 
with $\delta\phi=0.3\bar{\phi}$ (but not at $\delta\phi=0$) as the solution of 
(\ref{lambda}). 
Recall that the spinon Bose condensation stems from the equation (\ref{lambda}) 
with a nonzero $n_{BC}^b$ representing the density of spinons staying at $E_m=0$ 
state. In this case, there would be no solution at low temperature unless 
$\lambda$ takes a value to make $E_m$ gapless such that $n_{BC}^b\neq 0$
can balance the difference between the left and right side of the equation, 
similar to the half-filled case.\cite{bc} In particular, such a 
Bose condensation is found to be
sustained up to a finite temperature $T_{BC}\sim 0.21 J$ even in the present 2D 
case. This is due to the vanishingly small weight near $E=0$ in the density of states (in Fig. 4, there is a small tail in the solid curve which 
extends to $E=0$), where the spinon excitations at low-temperature are not 
sufficient to destroy the Bose condensation as first pointed out in Ref. 
\onlinecite{weng2}. It is noted that in principle the strength $\delta \phi$  of
the random fluctuations of $\delta A_{ij}^h$ should be self-consistently
determined by the density fluctuations of holons. But here for simplicity we
just treat $\delta \phi$ as a parameter and then study the qualitative characteristics under different values of $\delta \phi$. The actual strength
of the fluctuations of $A_{ij}^h$ will be only crucial in determining the 
location of the phase boundary.

{\it Phase separation} Note that the Bose condensation means a thermodynamic 
number of spinons staying at $E_m=0$ --- the lowest energy state which 
corresponds to the band edges of the spectrum $\bar{\xi}_m$. So due to 
the Bose condensation, such quantum states will acquire a macroscopic meaning.
But these band edge states of $\xi_m$ are very sensitive to the fluctuations
of $A_{ij}^h$ and the density of holons.
Physically, the density of holons is fluctuating in real space (which is the 
reason leading to the fluctuations of $A_{ij}^h$) and there
always exist those configurations in which the density of holes are relatively 
dilute in some areas where the flux described by $A_{ij}^h$ is reduced such that the band edge energies of $\xi_m$ can be close to $\pm 2J_s$. Since the probability would be small for such kind of 
inhomogeneous hole configurations, the density of states generally looks like a 
tail --- i.e., the Lifshitz tail--- near the band edges. Hence, the 
corresponding Bose-condensed state will generally look like having a charge 
inhomogeneity, or, phase separation, with spinons condensing into hole-deficient 
regions to form short-range spin ordering. The true AFLRO is absent here.

{\it Pseudo-gap behavior} The Bose condensation of spinons will also lead to a
pseudo-gap phenomenon. In magnetic aspect, for example, the density of
states shown in Fig. 4 indicates the suppression of spinon density of states 
between zero and the lowest peak, which stabilizes the Bose condensation as 
mentioned earlier. Since in the Bose condensation case there must be some
residual density extending to $E_m=0$, a pseudo spin gap is thus present in
$\rho_s(E)$. Its effect in the dynamic spin susceptibility will be discussed
later. In the transport aspect, holons which are the charge carriers
are scattered off by the gauge field $A_{ij}^s$ according to the holon effective
Hamiltonian (\ref{hh}). Anomalies in transport properties have been found in 
Ref. \onlinecite{weng2} with interesting experimental features in the similar 
effective Hamiltonian, where fluctuating fluxes depicted by 
$A_{ij}^s$ play a central role. But the Bose condensation of spinons will lead 
to a substantial suppression of $A_{ij}^s$ and thus a reduction of scattering to 
holons. Hence, the Bose condensation also provides an explanation for 
the  so-called pseudo-gap phenomenon shown in the {\it underdoped} high-$T_c$ 
cuprates, where the transport properties deviate from the 
high temperature ones below some characteristic temperature scale.

Therefore, if the Bose condensation of spinons happens in metallic phase, it
will result in a phase-separation or spin pseudo-gap phase without the
AFLRO. With the increase of $\delta$, the reduction of the left hand side of 
(\ref{lambda}) will eventually make the Bose condensation term, if exits,
disappear from the right hand side. So the Bose condensation in general may only
exist at small doping regime, which can be defined as the ``underdoping'' 
regime. In Fig. 3, the shaded curve sketches such a region outside the true
AFLRO phase which is in a much narrower region (dotted curve) at finite doping.

\subsection{Superconductivity}

In the phase-string representation, the operator of superconducting order parameter 
\begin{equation}
\hat{\Delta}^{SC}_{ij}\equiv \sum_{\sigma}\sigma c_{i\sigma}c_{j-\sigma}
\end{equation}
can be expressed in terms of (\ref{emutual}) as follows
\begin{equation}\label{pair}
\hat{\Delta}^{SC}_{ij}=\hat{\Delta}^s_{ij} 
\left(h_i^{\dagger}e^{\frac{i}{2}\Phi_i^b}\right)\left(h^{\dagger}_je^{\frac{i}
{2}\Phi^b_j} \right) (-1)^i,
\end{equation}
in which
\begin{equation}\label{spair}
\hat{\Delta}^s_{ij}\equiv \sum_{\sigma}e^{-
i\sigma A_{ij}^h}b_{i\sigma}b_{j-\sigma}.
\end{equation}
This is the basic expression to be used in the following discussion of 
superconducting condensation. 

\subsubsection{Mechanism}

For the sake of simplicity, we will focus on the nearest-neighboring pairing 
with $i=nn(j)$ below. Since the whole mean-field phase is built on 
\begin{equation} 
\langle \hat{\Delta}_{ij}^{s}\rangle={\Delta^s}\neq 0,
\end{equation}
the electron pairing order parameter g
$\Delta^{SC}_{ij}=\langle\hat{\Delta}^{SC}_{ij}\rangle$ can be 
written as
\begin{equation}\label{sc}
\Delta^{SC}_{ij}={\Delta^s} \left\langle \left(h_i^{\dagger}e^{\frac i 2
\Phi^b_i }\right)\left(h^{\dagger}_je^{\frac i 2 \Phi^b_j}\right)\right\rangle
(-1)^i.
\end{equation}
We see that the spinons are always paired in the present phase, as described by 
$\Delta^s$, up to a temperature scale $\sim J$ at small doping. Thus, in order
to have a real superconducting condensation below a 
transition temperature $T_c$, the holon part has to undergo a Bose condensation
or, strictly speaking in 2D, a superfluid transition in the Kosterlitz-Thouless 
sense (recall that both spinon and holon are {\it bosonic} in the present 
representation). 
 
One may notice that this superconducting condensation picture is
somewhat similar to that in the slave-boson mean-filed theory.\cite{gauge2} But 
there are two crucial differences. 

Firstly, the spinon pairing in the present case practically 
covers the whole superconducting and normal-state regime that we are interested 
in. In other words, $\Delta^s\neq 0$ in the present mean-field theory defines
a ``strange'' metal, and the normal-state anomalies of experimental measurements 
in the cuprates, including the magnetic properties and the transport properties, 
are all supposed to happen within such a phase. In contrast, in the slave-boson 
mean-field approach the fermionic spinon pairing is directly related to the {\it 
gap} in the spinon spectrum and has to {\it disappear} 
at a much lower temperature scale beyond which ``strange'' metallic properties 
presumably {\it start} to show up. 

Secondly, the Kosterlitz-Thouless transition temperature
of holons are believed to be much higher than the real $T_c$ in the cuprates, 
and thus one has to introduce other mechanism (e.g., gauge-field 
fluctuations\cite{gauge2}) to bring down the temperature scale in the 
slave-boson mean-field approximation. On the
other hand, there is a unique feature in (\ref{sc}), namely, the presence
of phases like $e^{\frac i 2 \Phi^b_i}$. Here $\Phi_i^b$ represents a structure
of vortices (anti-vortices) centered at $\uparrow$ ($\downarrow$) spinons. At
$T=0$, {\it all } $\uparrow$ and $\downarrow$ spinons
are paired up at a finite length scale and so are the vortices and 
anti-vortices in 
$\Phi^b_i$, which implies $\Delta_{ij}^{SC}\neq 0$ as long as holons are 
Bose-condensed.  At finite temperature, even though the Kosterlitz-Thouless 
transition temperature for hard-core bosons can be much higher, 
the ``phase coherence'' 
in $\Delta^{SC}_{ij}$ can be more quickly destroyed at a lower temperature due
to the dissolution of the vortices and antivortice bindings in $\Phi^b_i$ after 
{\it free} spinons appear. Here the argument for $\langle e^{\frac i 2 
\Phi^b_i}e^{\frac i 2 \Phi^b_j}\rangle=0$ is similar to the previous one for 
$\langle e^{i\Phi^h_i}\rangle=0$ which corresponds the disappearance of the 
long-range AF order once holons become mobile in the metallic phase. This 
provides us an estimate of the upper limit for the superconducting transition
temperature $T_c$ below.

\subsubsection{An estimate of $T_c$}

The holon effective Hamiltonian $H_h$ in  (\ref{hh}) determines the interaction 
between holons and those vortices described by $A_{ij}^h$. If free vortices are 
few, the condensed holons may easily ``screen'' them by forming supercurrent, 
which will then
effectively keep $\Delta^{SC}_{ij}$ finite. But if the number of free vortices,
or excited spinons, becomes comparable 
to the number of holons themselves, one expects that the ``screening'' effect 
collapses and thus $\Delta^{SC}_{ij}=0$. It predicts that $T_c$  will be 
basically determined by the spinon energy scale in the following way:
\begin{equation}\label{tc}
\left.2\sum_{m\neq 0}\frac{\lambda_m}{E_m}n(E_m)\right|_{T=T_c}=\kappa N\delta,
\end{equation}
where $\kappa\sim 1$ and the left-hand side represent the average number of 
excited spinons in
$\sum_{i\sigma} b^{\dagger}_{i\sigma}b_{i\sigma}$ with the Bose-condensed part 
(if exists) excluded. The dashed line in Fig. 3 represents
the $T_c$'s determined by (\ref{tc}) in the limit $\delta\phi\rightarrow 0$.
This curve may be regarded as represents the ``optimized'' $T_c$, because with
introducing the flux fluctuations, $\delta\phi\neq 0$, there is a finite 
density of states of spinons emerging at lower energy  which effectively 
reduces $T_c$
defined in (\ref{tc}). In the ``optimized'' limit of $\delta \phi=0$, one
may further simplify Eq.(\ref{tc}) by only retaining the contribution from the 
lowest-peak (which has a degeneracy $\delta N/2$) and obtains:
\begin{equation}\label{tc1}
T_c=\frac 1 c E_s,
\end{equation}
where $c$ is given by
\begin{equation}
c={\ln \left(1+\frac {2}{\kappa}\sqrt{1+(\xi_s/E_s)^2}\right)}>1.
\end{equation}
Here $E_s$ and $\xi_s$ are the energies of $E_m$ and $\xi_m$, respectively,
corresponding to the lowest-energy peak shown in Fig. 4 at $\delta\phi=0$.
Therefore, $T_c$ is indeed determined by the characteristic energy $E_s$ of 
spinon excitations.

\subsubsection{d-wave symmetry of the order parameter: phase string effect}

Finally, let us briefly discuss the symmetry of the order parameter 
$\Delta^{SC}_{ij}$. Basically, one needs to compare the relative phase of
$\Delta^{SC}_{ij}$ between $j=i+\hat{x}$ and $j=i+\hat{y}$, or
the phase change of the quantity $\tilde{h}_j^{\dagger}\equiv h_j^{\dagger} 
e^{\frac i 2 \Phi^b_j}$ in (\ref{sc}). Imagine that we move a holon from 
$j=i+\hat{x}$ to $i+\hat{y}$ via site $i+\hat{x}+\hat{y}$. At each 
step the holon has to exchange positions with a spinon with index $\sigma_{j'}$ 
at site $j'$ which leads to an extra phase $\sigma_{j'}=\pm 1$ due to $e^{\frac 
i 2\Phi^b_j}$ in $\tilde{h}_j^{\dagger}$. Even though other spinons
outside the path also contribute to, say, $\Phi^b_{i+\hat{x}+\hat{y}}-\Phi^b_{
i+\hat{x}}$, but their effect is canceled out as $h_j$ picks up the same phase change but with opposite sign in $H_h$. Therefore, in the end 
$\tilde{h}_j^{\dagger}$ acquires a total phase $\sigma_{i+\hat{y}}\cdot \sigma_{i+\hat{x}+\hat{y}}$ which is
just the {\it phase string} on such a path. Its contribution is
always {\it negative} on average for a short-ranged AF state. Assuming that 
this is the dominant path, one then concludes that $\Delta^{SC}_{ij}$ has to 
change sign from $j=i+\hat{x}$ to $i+\hat{y}$, namely, the d-wave symmetry. If 
only the nearest-neighbor-site electron pairing is considered, the order 
parameter in the momentum space can be written in the form $\Delta^{SC}({\bf 
k})\propto (\cos k_xa -\cos k_ya)$.  Therefore, 
the nonrepairable phase string effect and AF correlations are directly 
responsible for the d-wave symmetry of the superconducting condensation in the 
background of $\Delta^s\neq 0$. 

\subsection{Experimental implications: dynamic spin susceptibility }

\subsubsection{Local spin dynamic susceptibility}
 
The dynamic spin susceptibility function $\chi_L''(\omega)=1/N\sum_i 
\chi_{zz}''(i,i;\omega)$ describes the 
on-site spin dynamics and is derived in Appendix as follows 
\begin{eqnarray}\label{chil}
\chi_L''(\omega) & =& \frac {\pi} N \mathop{{\sum}'}_{mm'}\left(\sum_{i\sigma}
|w_{m\sigma}(i)|^2|w_{m'\sigma}(i)|^2\right) \nonumber\\
& & \cdot \left[ \frac{sgn(\omega)}{2}(1+n(E_m)+n(E_m'))(u_m^2v_{m'}^2+
v_m^2u_{m'}^2)\delta(|\omega|-E_m-E_{m'})\right. \nonumber\\
& & \left.+(n(E_m)-n(E_{m'}))(u_m^2u_{m'}^2+v^2_{m}v_{m'}^2)\delta(\omega+E_m-
E_{m'})\right],
\end{eqnarray}
where the summation $\sum'$ only runs over those $m$'s with $\xi_m<0$ (note that 
$E_m$ is symmetric under $\xi_m\rightarrow -\xi_m$).
If there is a Bose condensation of spinons, the contribution from the condensed 
part to $\chi''_L$ may be explicitly sorted out as
\begin{equation}\label{chilc}
\chi_c''(\omega)=sgn(\omega)\left(\frac{\pi}{2}n_{BC}^b\right)\frac{1}{N}\mathop{{\sum}'}_m K_{0m}\frac{\lambda_m}{E_m}\delta(\omega-E_m),
\end{equation} 
with $K_{0m}\equiv N\sum_{i\sigma}|w_{0\sigma}(i)|^2|w_{m\sigma}(i)|^2$ where 
the subscript $0$ refers to the $E_m=0$ state. 

Based on Eq.(\ref{chil}), two kinds of mean-field solutions, with and without 
Bose condensation of spinons, will be studied below. Without loss of generality, 
we consider these two cases at $\delta=0.143$ in which the corresponding
density of states, $\rho_s(\omega)$, is already shown in Fig. 4 for both cases.   
Let us firstly focus on the lowest peak of $\rho_s(E)$ shown in Fig. 4.
The contribution of such a peak to $\chi_L''(\omega)$ was previously discussed
in Ref.\onlinecite{letter} and is illustrated in Fig. 5 by the lowest 
sharp peak (dashed line) for $\delta\phi=0$ and the lowest twin peaks (solid curve) for $\delta\phi=0.3\bar{\phi}$, respectively. One sees two very distinct 
features here.  For the case of $\delta\phi=0.3\bar{\phi}$, there is a Bose 
condensation contribution 
at $T< T_{BC}\sim 0.21 J$ and it leads to a double-peak structure. But in 
$\delta\phi=0$ case, the Bose condensation is absent and one finds only a single 
sharp peak at $2E_s\sim 0.4J$. In other words, $\chi''_L(\omega)$ has 
drastically different characteristics for cases with and without a spinon 
Bose condensation. 

The twin-peak splitting may be understood as follows. The second 
peak corresponds a pair of spinons excited from the RVB vacuum,
while the first peak describes a {\it single} spinon excitation as the
other branch of spinon is in the Bose-condensate state. Such a lowest-peak basically maps out the lowest peak of the spinon density of states 
$\rho_s(E)$ in Fig. 4 according to (\ref{chilc}), and the second peak in 
$\chi''_L(\omega)$ is
located at an energy approximately twice larger than the first one. The latter 
one will be always around at low temperature no matter whether there is 
a Bose condensation or not. So there is a distinct behavior of those 
two peaks at different temperature as shown in Ref.\onlinecite{letter}, where
the weight of the lowest one gradually diminishes as the temperature approaches 
to $T_{BC}$. 

In contrast, there is only one sharp ``resonance-like'' peak left in the
where the Bose condensation is absent. It corresponds to a pair 
of spinon excitations located at the 
lowest peak ($E_s\sim 0.2J$) of $\rho_s$ in Fig. 4 with $\delta \phi=0$ (dashed 
curve). Note that this is the limit where the flux fluctuating part $\delta 
A_{ij}^h$ is totally suppressed such that a real spinon gap is opened
up at low-energy as shown in Fig. 4. The location of the ``resonance'' peak,
$2E_s$, is slightly lowered in energy as compared to the corresponding (second) 
peak in the twin-peak case at $\delta\phi=0.3\bar{\phi}$. The energy scale of 
this peak 
($2E_s\sim 0.4 J$) at $\delta\phi\rightarrow 0$ limit is roughly independent of 
$J_h$ and thus of $t$. This is because $J_h$ term only shifts $\lambda$
to $\lambda_m$ by a {\it constant} if there is no dispersion in $\xi_m$ near the
lowest (highest) peak, which will not affect $E_m$ near the lowest peak as
$\lambda$ will readjust its value correspondingly.

\subsubsection{Underdoping vs. optimal-doping}

We have previously discussed the Bose condensation of spinons and argued that it 
exists only in an ``underdoped'' regime. We have also shown that
holons as bosons can experience a Bose condensation at 
$T_c$, leading to the superconducting condensation. If the Bose condensation
temperature for holons is higher than $T_{BC}$ for spinons, i.e., 
$T_c>T_{BC}$, the holons will become Bose-condensed before spinons do and it
will be generally a {\it uniform} state since there is no inhomogeneous spin 
ordering above $T_{BC}$. Thus one may have $A_{ij}^h\approx\bar{A}_{ij}^h$ 
with $\delta\phi$ being much less than in the normal state. In this case, the
Bose condensation of spinons can be effectively prevented at lower temperature
because a homogeneous $\bar{A}_{ij}^h$ generally leads to an opening of a
real spinon gap as shown in Fig. 4 for $\phi=0$. So to be self-consistent, 
once $T_c>T_{BC}$, $T_{BC}$
may no longer exist and $T_c$ becomes the only meaningful temperature scale. 
Furthermore, we have already seen that $T_c$ is also optimized under 
$\delta\phi\sim 0$. Therefore, this region may be properly defined as
the ``optimal-doping'' regime in our theory in contrast to the 
previously defined underdoped regime at $T_{BC}>T_c$. 

Such an optimal-doping phase is charge
homogeneous and the Bose condensation of spinons is absent. It is
characterized, below $T_c$, by the ``resonance-like'' peak emerging in 
$\chi_L''(\omega)$ at $2E_s$ in Fig. 5. It is in accord with the $41$$meV$
peak found\cite{o7} in the optimally doped $YBa_2Cu_3O_7$ below $T_c$, if one chooses $J=100$$meV$ here. Above $T_c$, the ``resonance'' peak will quickly
disappear as the motion of holons becomes incoherent and a different behavior
of $A_{ij}^h$ is involved as discussed in Sec. II.

On the other hand, the underdoping regime with $T_{BC}>T_c$ is characterized by 
a low-energy twin-peak structure in $\chi_L''(\omega)$ at $T<T_{BC}$. In 
contrast to the ``optimal-doping'' case, such
an energy structure may not be qualitatively changed even when $T$ is below $T_c$, as holons are also expected to be condensed {\it inhomogeneously} 
in favor of the spinon energy. A twin-peak feature has been  
observed recently in the underdoped $YBa_2Cu_3O_{6.5}$ compound by neutron 
scattering\cite{o6.5} in the odd symmetry channel. In the experiment, the lowest
peak is located near $30 $$meV$ and the second one is near $60$$meV$, indeed 
about twice bigger in energy. 
Most recently, in the underdoped $YBa_2Cu_3O_{6.6}$, a second energy scale near 
$\sim 70 $$meV$ has been also indicated\cite{hayden1}
besides the earlier report of the lower energy peak near $34$$meV$.\cite{dai1}
It is noted that the energy scales shown in Fig. 5 are generally doping 
dependent, and those energies in $YBa_2Cu_3O_{6.5}$ are expected to be 
relatively smaller than the corresponding peaks in $YBa_2Cu_3O_{6.6}$. 

A word of caution about the comparison with the $YBCO$ compound is that the
latter is a double-layer system where two adjacent layer 
coupling is also important. But we do not expect the double-layer coupling 
to qualitatively change the above energy structure of  $\chi_L''$ in the odd 
symmetry channel. We point out that the
fluctuating part of the gauge-field $\delta A_{ij}^h$ usually makes the two 
adjacent layers difficult to couple together unless there are AF spin 
domains in the charge deficient region where the total $A_{ij}^h$ is suppressed,
as may be the case of phase separation. But in the uniform phase, with $\delta 
A_{ij}^h$ being suppressed below $T_c$ due to the Bose condensation of holons, the effective coupling between layers can also be greatly 
enhanced to gain interlayer-coupling spin energy. The Anderson's
confinement-deconfinement phenomenon\cite{book} may become most prominently in
the optimal-doping regime, which needs to be further explored. 

\subsubsection{Prediction}

Fig. 5 also shows $\chi_L''(\omega)$ in the whole energy regime at $\delta\phi/\bar{\phi}=0$ (dashed curve) and $0.3$ (solid curve), respectively.
As compared to the $\delta=0$ curve in the insert, a multi-peak structure 
is present as well at high energies for such a doped case. 
For example, if the $41$$meV$ peak in $YBa_2Cu_3O_7$ is explained
by the lowest peak in the case of $\delta\phi=0$, then the theory predicts
a second ``resonance'' peak near $120$$meV$ to be found. Those high energy peaks 
in Fig. 5
become rather closer in energy especially in the Bose condensed case (solid 
curve) which could be very easily smeared out either by the experimental 
solution (it may be further complicated by the fact that the momentum-dependence 
varies drastically among those peaks) or by the dynamic broadening due to the finite life time of spinons which is beyond the present mean-field treatment. Nevertheless, the multi-peak structure, especially the twin-peak
feature at low energy in the spinon Bose-condensed case, should become 
observable by high solution measurement at low temperature as the unique
prediction of the present theory. 

\section{DISCUSSIONS}

In this paper, we have approached the doped antiferromagnet from the 
half-filling side, where the bosonic RVB description is known
to be very accurate for the antiferromagnetism. The crucial modification at 
finite doping comes from the phase string effect induced by doped 
holes. For example, doped holes are turned into bosonic holons by
such a phase string effect so that both the elementary spin and charge 
excitations are bosonic. The Bose condensation of spinons in insulating phase
and the Bose condensation of holons in metallic phase determine the AFLRO and
superconducting phase transitions, respectively. 

While the bosonic RVB pairing, representing short-range AF 
correlations, is always present and is the driving force behind the 
antiferromagnetism and superconductivity, it is the combination with the phase 
string effect that 
decides when and where they occur in the phase diagram. For instance, the 
Bose-condensation of spinons leads to the AFLRO only in the case that holes are localized. In the metallic phase where holes become mobile, the AFLRO will
be destroyed by the phase string effect. But the Bose-condensation of 
spinons may still persist into weakly doped metallic region, leading to an
``underdoping'' metallic phase with charge inhomogeneity (phase separation) and pseudo-gap phenomenon. 

There still are many theoretical and experimental issues 
which have not been dealt with in the present paper and are left for further
investigation. Here we conclude by giving several critical remarks.
The first is about the phase diagram at larger doping. Recall that in our
mean-field description, the metallic phases are characterized by two temperature
scales: $T_c$ and $T_{BC}$, and we have argued that $T_c>T_{BC}$ determines an
``optimal-doping'' regime at low temperature where $T_{BC}$ is no longer 
meaningful. But beyond this regime, there is a possibility that holons may tend 
to be always Bose-condensed even at {\it normal state} in favor of the hopping 
energy. If this occurs, the gauge field $A_{ij}^f$ in $H_J$ may 
have to be ``expelled'' to the spinon part, leading to a statistics 
transmutation to turn spinons into {\it fermions} and causing a collapse of the 
bosonic
RVB order parameter at the normal state. In this picture, the normal state in overdoped regime may simply recover the fermionic uniform RVB 
state.\cite{gauge2}       

The second issue is about the time-reversal symmetry. Recall that the sharp peak 
structure in spinon spectrum below $T_c$ is mathematically similar to a Landau
level structure in a uniform magnetic field. One may naturally wonder if
some kind of time-reversal symmetry would be apparently broken like in the anyon 
theories.\cite{laughlin} Below we point out that this is not the case in the 
present theory. First of all, it is easy to see that there is no 
breaking of the time-reversal symmetry  
in the holon Hamiltonian $H_h$ (\ref{hh}) in which the gauge field 
$A_{ij}^f=A^s_{ij}-\phi^0_{ij}$. Here $A_{ij}^s$ behaves like a fluctuating 
gauge field with 
$\langle A^s_{ij}\rangle=0$, and $\phi^0_{ij}$ describes a uniform $\pi$-flux 
per plaquette which does not break the time-reversal symmetry either as a gauge 
transformation can easily 
change $\pi$ flux into $-\pi$ flux per plaquette. As for the spin part, even 
though spinons see $A_{ij}^h$ which breaks the time-reversal symmetry, one
should remember that the physical 
observable quantity is the spin-spin correlation functions like $\chi_L''$ shown
in (\ref{chil}), which can be easily shown to be invariant under 
$A_{ij}^h\rightarrow -A^h_{ij}$.

Lastly, the sharp peak in the spinon spectrum in the uniform phase 
provides an explanation for the $41$$meV$ peak in the neutron-scattering measurement of 
$YBaCuO$ ``90 K'' sample, but it also means a {\it real} gap in the spinon 
spectrum. From a naive spin-charge separation picture, 
one would expect the single-electron Green's function to be a convolution of
spinon and holon propagators and the electron spectrum may also show a finite
gap as well in the superconducting state which may be inconsistent with d-wave 
symmetry. But 
we note that in the present spin-charge separation formalism (\ref{emutual}),
there is an additional phase field $\Theta_{i\sigma}^{sting}$ representing the 
phase string effect. It means that if a ``bare'' hole created by $c_{i\sigma}$ 
decays into mobile spinon and holon, a nonlocal topological effect will be left 
behind which could cost a logarithmic-divergent energy. In other words, the 
phase string effect in 2D case will serve as a confinement force to prevent a 
newly doped hole from dissolving into elementary excitations. (Of course,
internal charge and spin excitations without involving the change of total 
electron number 
are still described by the spin-charge separation in the present mean-field 
theory.) In fact, even in 1D case, the
single-electron Green's function looks quite differently from a simple 
convolution of spinon and holon propagators, and recently 
Suzuura and Nagaosa\cite{suzuura} have discussed the crucial role of the
phase string effect in understanding the angle-resolved photoemission 
spectroscopy in $SrCuO_2$.\cite{kim} Our preliminary investigation in 2D 
indicates that a bare 
``hole'' wavepacket injected into the background of the spin-charge
separation mean-field state will behave more like a conventional band-structure 
quasiparticle in a Fermi liquid which shows d-wave gap structure
when the holons are Bose-condensed and pseudo-gap structure when spinons are
Bose-condensed. So experiments involving injecting electron or hole into the 
system like photoemission spectroscopy may no longer provide 
{\it direct} information of elementary excitations like in the conventional 
Fermi liquid theory, due to the confinement of phase string effect.  

\acknowledgments

We have benefited from helpful discussions with Yong-Cong Chen and T. K. 
Lee, H. F. Fong, and P. Bourges. The present work is supported by the grants 
from the Texas ARP No. 3652707 and Robert A. Welch foundation and, and
by Texas Center for Superconductivity at University of Houston.\\ 

\appendix{\section{Dynamic spin susceptibility function}}

Local spin susceptibility function is defined in the Matsubara 
representation as follows
\begin{equation}
\chi_{\alpha \beta}(i,i; i\omega_n)=\int^{\beta}_0d{\tau}\ \ e^{i\omega_n\tau}\langle T_{\tau} S^{\alpha}_i(\tau)
S^{\beta}_i(0)\rangle,
\end{equation}
where $\omega_n=\frac{2\pi n}{\beta}$. In the following we will determine
the dynamic spin susceptibility function $\chi_{\alpha\beta} ''(i,i;\omega)$
based on the present mean-field theory.

Consider $\langle T_{\tau} S^{\alpha}_i(\tau)
S^{\beta}_i(0)\rangle$ in $\alpha=\beta={z}$ case.
In the present mean-field formulation, one has
\begin{eqnarray}\label{c4}
\left\langle T_{\tau} S^z_i(\tau)S^z_i(0)\right\rangle 
& = &\frac 1 4 \sum_{\sigma\sigma'}\sigma\sigma'\left\langle 
T_{\tau}b^{\dagger}_{i\sigma} (\tau)b_{i\sigma}(\tau)b^{\dagger}_{i\sigma'}(0)
b_{i\sigma'}(0)\right\rangle \nonumber\\
& = &\frac 1 4 \sum_{\sigma}\left\langle
T_{\tau}b^{\dagger}_{i\sigma}(\tau)b_{i\sigma}(0)\right\rangle\left\langle 
T_{\tau}b_{i\sigma}(\tau)b^{\dagger}_{i\sigma}(0)\right\rangle.
\end{eqnarray}
In terms of (\ref{bogo2}) and (\ref{hs}), one has
\begin{eqnarray}
b^{\dagger}_{i\sigma}(\tau)& \equiv & e^{H_s\tau}b_{i\sigma}^{\dagger} e^{-
H_s\tau}\nonumber\\ &=&\sum_m\left(u_m\gamma_{m\sigma}^{\dagger} e^{E_m\tau}-v_m 
\gamma_{m-\sigma}e^{-E_m\tau}\right){ w}^*_{m\sigma}(i)e^{-
i\sigma\chi_m}.
\end{eqnarray}
Then by noting 
$w^*_{m-\sigma}=w_{m\sigma}$ and that for each $m$ with $\xi_m<0$, one always
can find a state $\bar{m}$ with $\xi_{\bar{m}}=-\xi_m>0$ with a wavefunction
\begin{equation}\label{wp3}
w_{\bar{m}\sigma}(i)=(-1)^iw_{m\sigma}(i)
\end{equation}
according to (\ref{ew}),
we get
\begin{eqnarray}
\langle T_{\tau} b^{\dagger}_{i\sigma}(\tau) b_{j\sigma}(0)\rangle_{\tau>0}
=2\mathop{{\sum}'}_m w^*_{m\sigma}(i)w_{m\sigma}(i)
\times \left[u_m^2 n(E_m) e^{E_m\tau}  + v_m^2 (1+n(E_m))e^{-E_m\tau}\right],
\end{eqnarray}
where the summation $\sum'_m$ only runs over those states with $\xi_m<0$.
Then it is straightforward to obtain $\chi_{zz}$ after integrate out $\tau$ in (\ref{c4}): 
\begin{eqnarray}\label{chizz}
\chi_{zz}(i,i; i\omega_n)= \chi_{zz}^{(-)}(i,i; i\omega_n)+ 
\chi_{zz}^{(+)}(i,i; i\omega_n),
\end{eqnarray}
where $\chi^{(\pm)}_{zz}$ is defined by
\begin{eqnarray}
{\chi_{zz}^{(\pm)}(i,i;i\omega_n)=\frac 1 2 \mathop{{\sum}'}_{mm'} K^{zz}_{mm'}(i,i)
\left[(p_{mm'}^{\pm})^2 \frac {n(E_{m'})-n(E_{m})}{i\omega_n+E_m-E_{m'}}\right. }\nonumber\\
\left. +(l_{mm'}^{\pm})^2 \left(1+n(E_m)+n(E_{m'})\right) \frac 1 2 \left(\frac 
1 {i\omega_n+E_m+E_{m'}}-\frac 1 {i\omega_n-E_m-E_{m'}}\right)\right] 
\end{eqnarray} 
with 
\begin{equation}\label{kzz}
K^{zz}_{mm'}(i,i)\equiv \sum_{\sigma}\left|w_{m\sigma}(i)\right|^2\left|w_{m'\sigma}(i)\right|^2.
\end{equation}
Here the coherent factors, 
$p^{\pm}_{mm'}$ and $l^{\pm}_{mm'}$ are defined by 
\begin{eqnarray}
p^{\pm}_{mm'}& = & u_mu_{m'}\pm v_mv_{m'}, \nonumber\\
l^{\pm}_{mm'}& = & u_mv_{m'}\pm v_mu_{m'}.
\end{eqnarray}
Finally, the dynamic spin susceptibility function $\chi_{zz}''(i,i; \omega)$ can 
be obtained as the imaginary part of $\chi_{zz}$ after an analytic  
continuation $i\omega_n\rightarrow \omega +i 0^+$ is made:
\begin{equation}\label{ichizz}
\chi''_{zz}(i,i;\omega)=\Phi_{zz}^{(-)}(i,i; \omega) +  
\Phi_{zz}^{(+)}(i,i; \omega),
\end{equation}
where 
\begin{eqnarray}\label{phi}
\lefteqn{\Phi_{zz}^{(\pm)}(i,i;\omega) 
=\frac{\pi}{4}\mathop{{\sum}'}_{mm'}K^{zz}_{mm'}(i,i)\left[\left(1+n(E_m)+n(E_{m'})
\right) (l_{mm'}^{\pm})^2 sgn (\omega)\right.} \nonumber\\
&\left.\cdot\delta(|\omega|-E_m-E_{m'})+ 2 \left(n(E_m)-
n(E_{m'})\right)(p_{mm'}^{\pm})^2\delta (\omega+E_m-E_{m'})\right].
\end{eqnarray}

\figure{Fig. 1 A sequence of sign mismatches (with reference to a spin 
background satisfying the Marshall sign rule) left by the hopping of a
hole from site $a$ to site $b$ on square lattice.\label{fig1}} 

\figure{Fig. 2 Fictitious $\pi$ flux-tubes bound to holons which can only be 
seen by spinons and are described by the gauge field $A_{ij}^h$ 
defined by (\ref{eah}).\label{fig2}}

\figure{Fig. 3 Phase diagram of the bosonic RVB state characterized by the
order parameter $\Delta^s$ (solid curve). Within this phase, the
shaded curve sketches a region where a Bose condensation of spinons (BC) occurs,
which leads to the AFLRO in an insulating phase (dotted curve) but a charge
inhomogeneity with short-ranged spin ordering in metallic region which defines
an underdoped regime. Superconducting condensation (SC)  
happens due to the Bose condensation of holons and $T_c$ (dashed curve) is 
determined under an optimal condition (see the 
text).\label{fig3}}

\figure{Fig. 4 Spinon density of states at $\delta=0.143$ and $T=0$. Solid curve 
corresponds to $\delta\phi=0.3\bar{\phi}$ and the dashed curve is for 
$\delta\phi=0$. Insert:
the density of states at half-filling. The energy is in units of $J$. 
\label{fig4}}

\figure{Fig. 5 Local dynamic spin susceptibility versus energy at 
$\delta=0.143$. Solid curve: $\delta\phi=0.3\bar{\phi}$ and dashed curve:
$\delta\phi=0$ at $T=0$. Note that the lowest peak of dashed curve splits into 
a twin-peak structure in solid curve (see text). The half-filled case is shown 
in the insert for comparison.\label{fig5}}

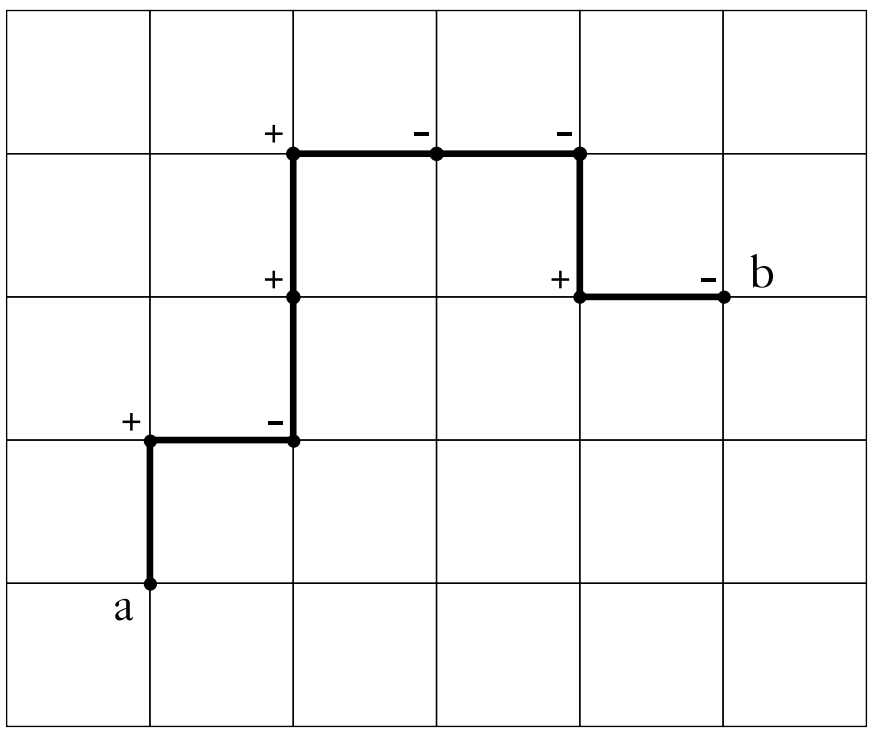
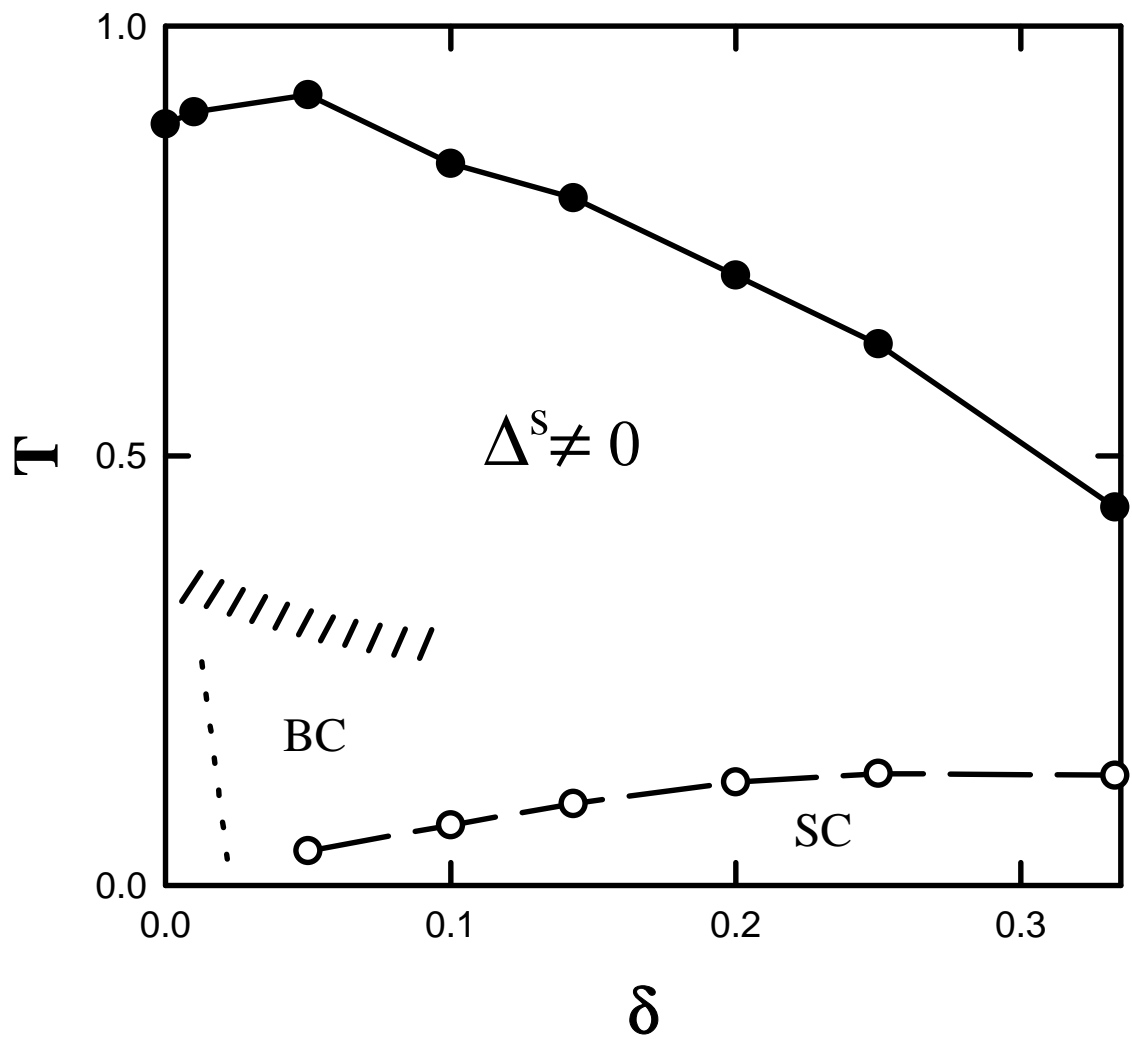
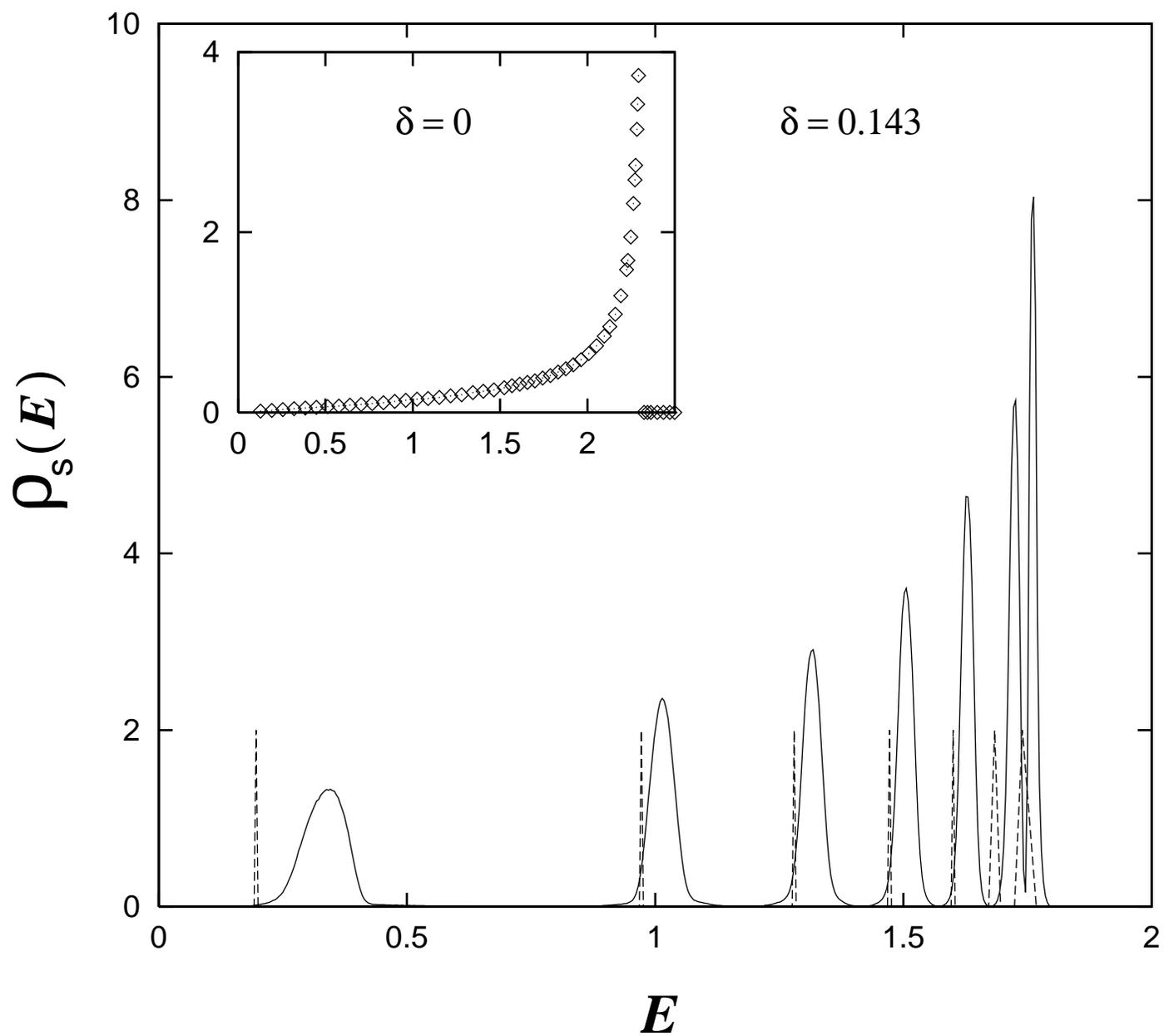
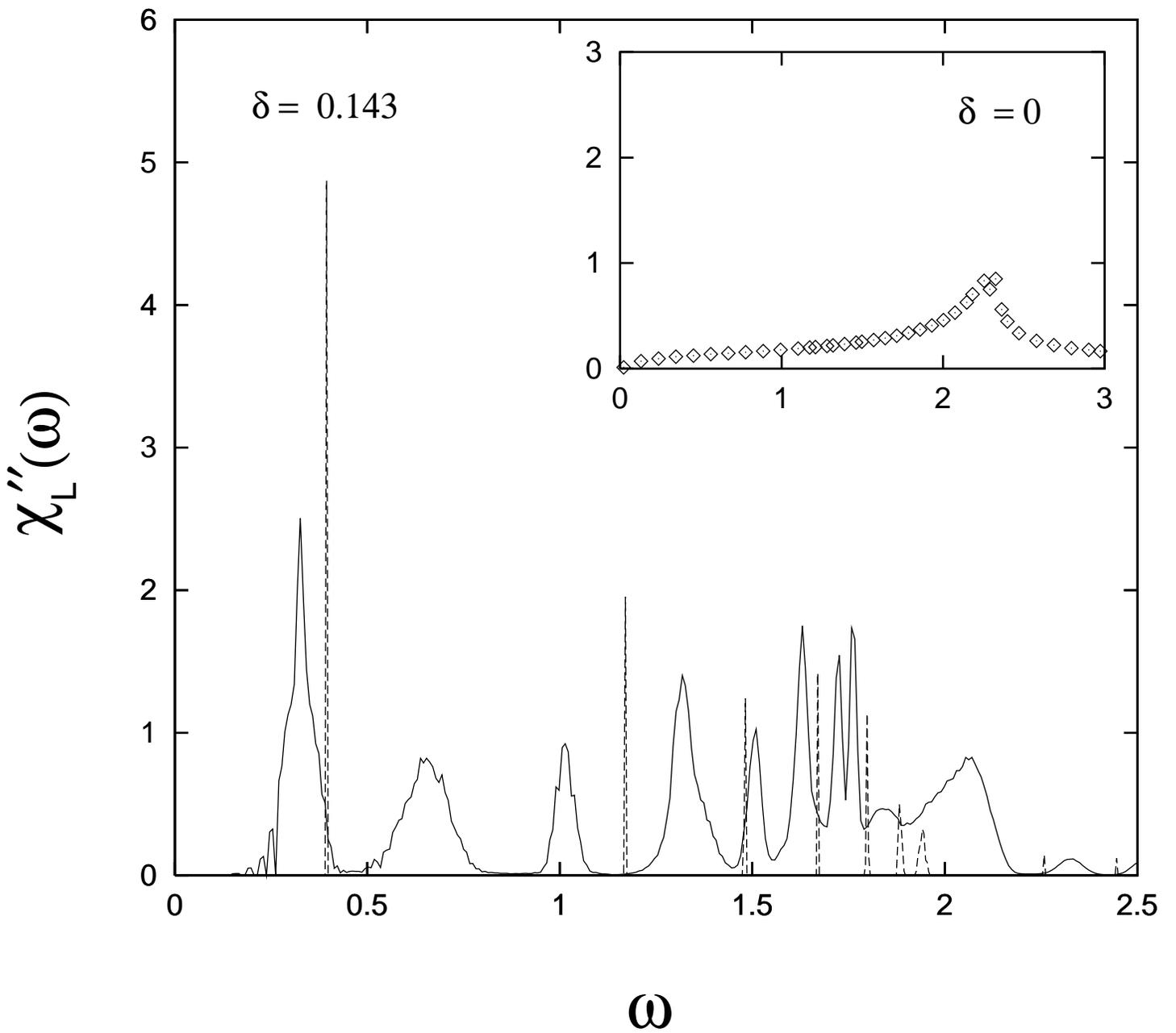
\end{document}